\documentclass[letterpaper, 10 pt, conference]{ieeeconf}  

\IEEEoverridecommandlockouts                              

\usepackage{caption}
\usepackage{graphics} 
\usepackage{amsmath} 
\usepackage{amssymb}  
\usepackage{mathtools}

\usepackage{amsthm}
\usepackage{bbold}
\usepackage{cite}
\usepackage{subcaption}
\usepackage{hyperref}
\usepackage{xcolor}
\usepackage{tikz}
\usepackage{circuitikz}
\usetikzlibrary{positioning}

\usepackage{balance}

\title{\LARGE \bf
Decentralized Continuification Control of Multi-Agent Systems via Distributed Density Estimation
}
\author{Beniamino Di Lorenzo$^{1}$, Gian Carlo Maffettone$^{1}$ and Mario di Bernardo$^{1, 2,*}$
\thanks{{This work was financially supported by the Italian Ministry of University and Research (MUR) under the PRIN 2022 project titled ’Machine-learning based control of complex multi-agent systems for search and rescue operations in natural disasters (MENTOR).}}
\thanks{$^{1}$Modeling and Engineering Risk and Complexity, Scuola Superiore Meridionale, Largo san Marcellino 10, 80136, Naples, Italy. (emails:
        {b.dilorenzo@ssmeridionale.it, giancarlo.maffettone@unina.it, mario.dibernardo@unina.it})}%
\thanks{$^{2}$Department of Electrical Engineering and Information Technology, Univeristy of Naples Federico II, via Claudio 21, 80125, Naples, Italy.}%
\thanks{$^*$ corresponding author}
}

\newtheorem{theorem}{Theorem}

\begin{document}

\maketitle
\thispagestyle{empty}
\pagestyle{empty}

\begin{abstract}
This paper introduces a novel decentralized implementation of a continuification-based strategy to control the density of large-scale multi-agent systems on the unit circle. While continuification methods effectively address micro-to-macro control problems by reformulating ordinary/stochastic differential equations (ODEs/SDEs) agent-based models into more tractable partial differential equations (PDEs), they traditionally require centralized knowledge of macroscopic state observables. We overcome this limitation by developing a distributed density estimation framework that combines kernel density estimation with PI consensus dynamics. Our approach enables agents to compute local density estimates and derive local control actions using only information from neighboring agents in a communication network. Numerical validations across multiple scenarios—including regulation, tracking, and time-varying communication topologies--confirm the effectiveness of the proposed approach. They also convincingly demonstrate that our decentralized implementation achieves performance comparable to centralized approaches while enhancing reliability and practical applicability. 
\end{abstract}

\section{INTRODUCTION} \label{sec:introduction}

The challenge of controlling large-scale multi-agent systems appears across diverse applications including transportation networks \cite{ferrara2018freeway}, biological systems \cite{bernoff2011primer}, and swarm robotics \cite{elamvazhuthi2019mean}. A key research question in this domain involves determining appropriate agent-level control interventions at the microscopic level that produce desired collective behaviors at the macroscopic level \cite{d2023controlling}. 
For example, in traffic management, an open crucial problem is to design the behaviour of relatively small sets of autonomous vehicles to influence the overall traffic flow, creating smoother density distributions \cite{liu2021learning}. Solving this micro-to-macro control problem represents a significant frontier in complex systems, requiring innovative approaches to bridge individual actions and collective outcomes.

\begin{figure}
     \centering
     \includegraphics[width=0.8\linewidth]{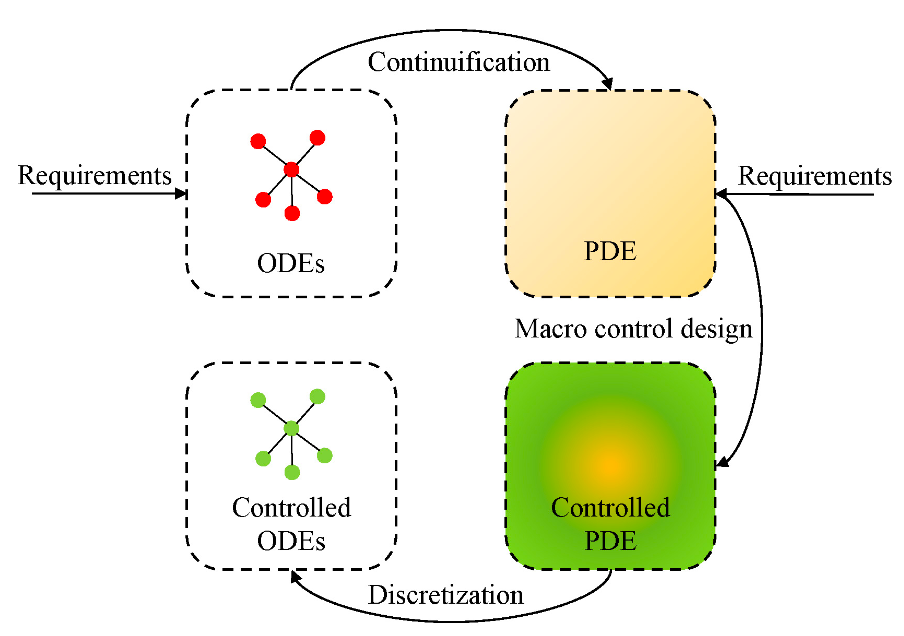}
     \caption{Continuification control pipeline, inspired by \cite{nikitin2021continuation}.}
     \label{fig:continuification}
 \end{figure}

From the methodological standpoint, macroscopic control approaches have recently gained momentum as alternatives to agent-based microscopic techniques \cite{gardi2022microrobot,gazi2011swarm}. In this context, models direclty describes the collective behavior to control instead of focusing on the individuals' behavior. This allows for scenarios with improved analytical tractability and scalability. 
Within this emerging field, continuification stands as a promising paradigmatic methodology \cite{nikitin2021continuation,maffettone2022continuification, maffettone2024mixed}.
This method provides a systematic pipeline for addressing multi-scale control problems. The core idea is to reformulate large sets of ordinary/stochastich differential equations (ODEs/SDEs) into smaller sets of partial differential equations (PDEs) describing the emergent behavior to be controlled. This more compact, macroscopic model facilitates the use of analytical methods for control design. Once the macroscopic control action has been designed, it then needs to be discretized into feasible control inputs for the agents in the group (see Fig. \ref{fig:continuification} for a schematics). 
This technique has been succesfully applied to swarm robotics problems \cite{maffettone2024mixed} and for the synchronization of spin-torque oscillators \cite{nikitin2023synchronization}.

When employing macroscopic methods such as continuification, a significant limitation is the inherently centralized nature of the derived control. These strategies depend on macroscopic state descriptors (e.g., density) that are assumed to be globally known and accessible.

Despite recent efforts to understand how limited sensing and perturbations affect solutions \cite{maffettone2023continuification,maffettone2024high}, continuification methods have not yet been implemented in a fully decentralized manner to the best of our knowledge. In this work, we address this gap by decentralizing continuification control through distributed density estimation techniques and consensus dynamics thereby enhancing its reliability.

Density estimation involves constructing probability density function estimates from observed data \cite{silverman2018density}. Parametric algorithms assume samples originate from a known parametric distribution family and focus on parameter estimation \cite{gu2008distributed}. The effectiveness of parametric techniques depends on the validity of the assumed model, making them unsuitable for time-varying density estimation. Non-parametric approaches, by contrast, avoid assumptions about the underlying distribution family. Among these, kernel density estimation (KDE) has emerged as a particularly effective solution \cite{chen2017tutorial}. In distributed filtering algorithms, each agent implements a local estimator using its observations while exchanging information with neighboring agents to progressively refine the global distribution estimate \cite{kia2019tutorial}.

In this paper, we propose a novel decentralized implementation of continuification control using kernel density estimation and PI consensus, a dynamic average consensus algorithm, to estimate densities in a fully decentralized manner. 
While our set-up focus on one-dimensional settings for simplicity, the approach can be readily extended to higher dimensions, by adopting an approach similar to the one described in \cite{maffettone2024mixed} for the centralized continuification strategy.

\section{Model and Problem Statement} \label{sec:model_and_problem}

As in \cite{maffettone2022continuification}, we consider $N$ identical agents interacting on the circular domain $\Omega=[-\pi,\pi]$ according to
\begin{equation} \label{eq:agent-based_model}
    \dot{x}_i = \sum_{j=1}^N f(\{ x_i,x_j \}_\pi) + u_i, \quad i=1,\dots,N.
\end{equation}
 where $x_i\in\Omega$ is the position of the $i$-th agent, $\{x_i,x_j\}_\pi$ is the relative position between agents wrapped to have values in $\Omega$, $u_i$ is a velocity control input and $f:\Omega\rightarrow\mathbb{R}$ is a velocity interaction kernel. Similar to \cite{bernoff2011primer}, $f$ is assumed to be a vanishing odd function, that can express different ranges of attraction and/or repulsion (see \cite{maffettone2022continuification} for a review of the open loop behaviors for different choices of $f$). 
 
Assuming a sufficiently large value of $N$, the agents' macroscopic spatial organization can be compactly captured by a density function, defined as $\rho:\Omega \times \mathbb{R}_{\geq 0}  \rightarrow \mathbb{R}_{\geq 0}$, with, $\int_\Omega \rho(x, t) \,\mathrm{d}x = N$ for all $t\in\mathbb{R}_{\geq 0}$.

We define $\hat{\rho}^{(i)}$ as an estimate of the group's density available to agent $i$, and we consider that agents can exchange information about such an estimate through a communication network that we describe as an undirected graph $\mathcal{G}(V, E)$, where $V=\{1, \dots, N\}$ is the set of vertices and $E = \{(i, j): i, j\in V, \forall i \neq j\}$ is the set of edges.

As in \cite{kia2019tutorial}, given the set of neighbors of agent $i$, $\mathcal{N}^{(i)} = \{j\in V : (i, j) \in E\}$, we model the dynamics of the density estimation as
\begin{equation} \label{eq:density_estimator}
    \hat{\rho}^{(i)}_t(x,t) = c\left(\hat{\rho}^{(i)}(x,t),\{\hat{\rho}^{(j)}(x,t)\}_{j\in\mathcal{N}^{(i)}}\right),
\end{equation}
where the subscript represents a partial time derivative and $c$ is some communication protocol that will be defined later.

\begin{figure}
    \centering
    \includegraphics[width=\linewidth]{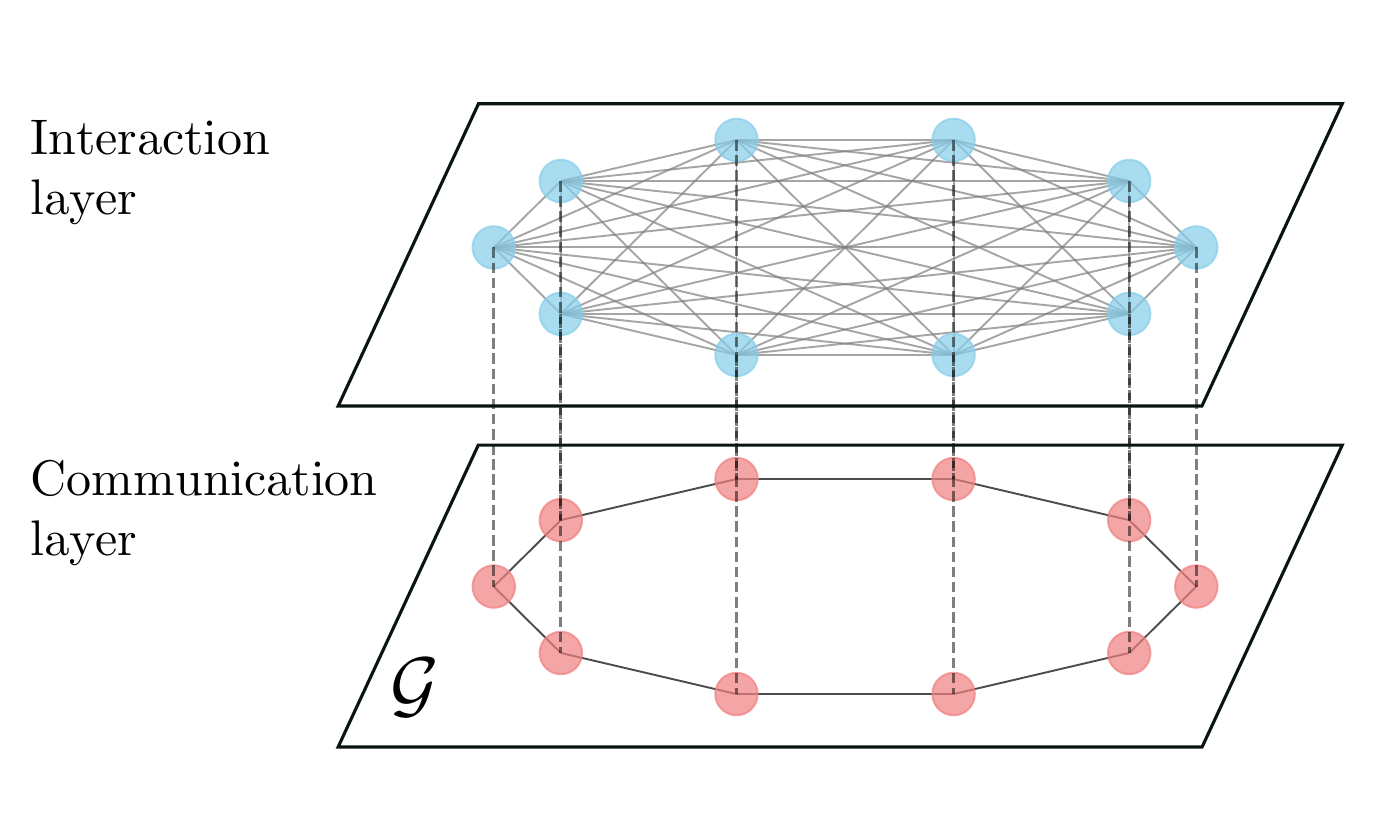}
    \caption{Schematization of agents interactions and communication. Agents physically influence each other through the interaction topology, and exchange information via the communication layer.}
    \label{fig:multilayer_network}
\end{figure}

Our set-up is analogous to agents interacting on a two-layer multiplex network as depicted in Fig. \ref{fig:multilayer_network}. The interaction layer is an all-to-all network (with the intensity of interactions decaying with the distance between agents, since $f$ is vanishing), while $\mathcal{G}$ describes the communication network, through which agents share information for density estimation. We assume $\mathcal{G}$ is strongly connected.

Our objective is to design distributed control inputs $u_i$ to make the agents arrange themselves according to some desired macroscopic configuration on $\Omega$. Given a desired density profile $\rho^\mathrm{d}$, the problem is to find $u_i$ in \eqref{eq:agent-based_model} such that 
\begin{equation} \label{eq:problem_statement}
    \lim_{t\to\infty} \| \rho^\mathrm{d}(\cdot,t) - \rho(\cdot,t) \|_2 = 0.
\end{equation}

We want agents to compute $u_i$ in a distributed manner using only local information. Therefore, we assume they have access only to the density estimate $\hat{\rho}^{(i)}$, evolving according to \eqref{eq:density_estimator}.
Specifically, we require $u_i$ to be solely a function of the $i$-th agent's density estimate:
\begin{equation} \label{eq:microscopic_inputs}
    u_i(t) = U^{(i)} \left(\rho^\mathrm{d}(x,t),\hat{\rho}^{(i)}(x,t)\right).
\end{equation}

Therefore, the communication protocol $c$ in \eqref{eq:density_estimator} must be designed to fulfill \eqref{eq:problem_statement}, despite the sensing constraints.

\section{CENTRALIZED CONTROL DESIGN} \label{sec:control_design}
We start by following the continuification approach described in \cite{maffettone2022continuification} and depicted in Fig. \ref{fig:continuification}. The strategy consists of the following steps: (i) lift the microscopic problem in \eqref{eq:agent-based_model} through a mean-field based continuification, returning a mass conservation law in space and time; (ii) perform macroscopic control design with convergence guarantees, and (iii) discretize the control action to obtain microscopic agents' inputs. 

We describe below in greater detail each of the steps above.

\subsection{Continuification}
Via mean-field theory \cite{bernoff2011primer, maffettone2022continuification}, when $N\to\infty$ we can recast \eqref{eq:agent-based_model} as the mass balance law
\begin{equation} \label{eq:macroscopic_model}
    \rho_t(x,t) + \left[\rho(x,t)V(x,t)\right]_x = q(x, t),
\end{equation}
where the velocity field $V$ captures microscopic all-to-all interactions in the continuum as
\begin{equation}
    V(x,t) = \int_{-\pi}^{\pi} f \left(\{x,y\}_\pi \right) \rho(y,t) \, \mathrm{d}y = (f * \rho) (x,t)
\end{equation}
with  ``$*$" denoting the circular convolution operator. The continuum control action is represented as
\begin{align}\label{eq:q_and_U}
    q(x, t) = -\left[\rho(x, t) U(x, t)\right]_x,
\end{align}
where the control field $U$ is the macroscopic counterpart of $u_i$ in \eqref{eq:agent-based_model}.

For the problem to be well posed, we impose periodic boundary conditions and initial conditions:
\begin{align}
    \rho(-\pi,t) &= \rho(\pi,t), \quad \forall t \geq 0, \\
    \rho(x,0) &= \rho^0(x), \quad \forall x \in [-\pi,\pi].
\end{align}
These periodic boundary conditions ensure that, when $q=0$, mass is conserved, i.e. $ (\int_{-\pi}^{\pi} \rho(x,t) \, \mathrm{d} x )_t=0$ (recall that $V$ is periodic by construction as it results from a circular convolution; see \cite{maffettone2022continuification} for more details). 

\subsection{Macroscopic control design}
We design $q$ assuming that the desired density profile satisfies the reference dynamics
\begin{equation}
    \rho^\mathrm{d}_t(x,t) + [\rho^\mathrm{d}(x,t) V^\mathrm{d}(x,t)]_x = 0,
\end{equation}
where
\begin{equation}
    V^\mathrm{d}(x,t) = \int_{-\pi}^{\pi} f(\{x,y\}_\pi)\rho^\mathrm{d}(y,t) \, \mathrm{d}y = (f * \rho^\mathrm{d}) (x,t).
\end{equation}
Defining the error function as
\begin{equation}
    e(x,t) = \rho^\mathrm{d}(x,t) - \rho(x,t).
\end{equation}
we can state the following theorem. 

\begin{theorem} [Macroscopic Convergence]\label{th:macro_convergence}
Choosing 
\begin{equation} \label{eq:source_term}
    q(x,t) = K_\mathrm{p} e(x,t) + [\rho(x,t) V(x,t)]_x - [\rho^\mathrm{d}(x,t) V^\mathrm{d}(x,t)]_x,
\end{equation}
where $K_\mathrm{p}$ is a positive control gain, the error dynamics globally asymptotically pointwisely converges to 0.
\end{theorem}
\begin{proof}
    See Theorem 1 in \cite{maffettone2024mixed} for $d = 1$.
\end{proof}

\subsection{Discretization}\label{subsec:discretization}
Next, we discretize the macroscopic control action to recover agent-level control inputs. 
Exploiting \eqref{eq:q_and_U}, we recover $U$ from $q$ through spatial integration, yielding 
\begin{equation} \label{eq:macroscopic_control_centralized}
    U(x,t) = - \frac{1}{{\rho}(x,t)} \left[ \int_{-\pi}^x {q}(y,t) \, \mathrm{d} y + q(-\pi,t)\right].
\end{equation}

Agents can then compute their individual control inputs $u_i$  through spatial sampling of \eqref{eq:macroscopic_control_centralized}, that is by setting $u_i(t) = U(x_i, t)$ \cite{maffettone2022continuification}.

\section{Decentralized continuification control}
Next, we assume that, in accordance with the constraints described in Section \ref{sec:model_and_problem}, agents cannot access $U$ directly, but must estimate it through communication with  their neighbors. Therefore, each agent in the group can only compute a local estimate of the macroscopic control action given by:
\begin{equation} \label{eq:macroscopic_control}
    U^{(i)}(x,t) = - \frac{1}{\hat{\rho}^{(i)}(x,t)} \left[ \int_{-\pi}^x \hat{q}^{(i)}(y,t) \, \mathrm{d} y + \hat{q}^{(i)}(-\pi,t)\right]
\end{equation}
where $\hat{\rho}^{(i)}$ and $\hat{q}^{(i)}$ are the agent's local estimates of $\rho$ and $q$.

The microscopic control input $u_i$ can then obtained through spatial sampling of $U^{(i)}$ as:
\begin{equation}
    u_i(t) = U^{(i)}(x_i,t), \quad \forall i=1,\dots, N.
\end{equation}

To close the loop, agents need to be able to estimate $\hat{\rho}^{(i)}$ and $\hat{q}^{(i)}$ as will be discussed in what follows.

\begin{figure} 
    \centering
    \subfloat[]{%
        \includegraphics[width=0.5\textwidth]{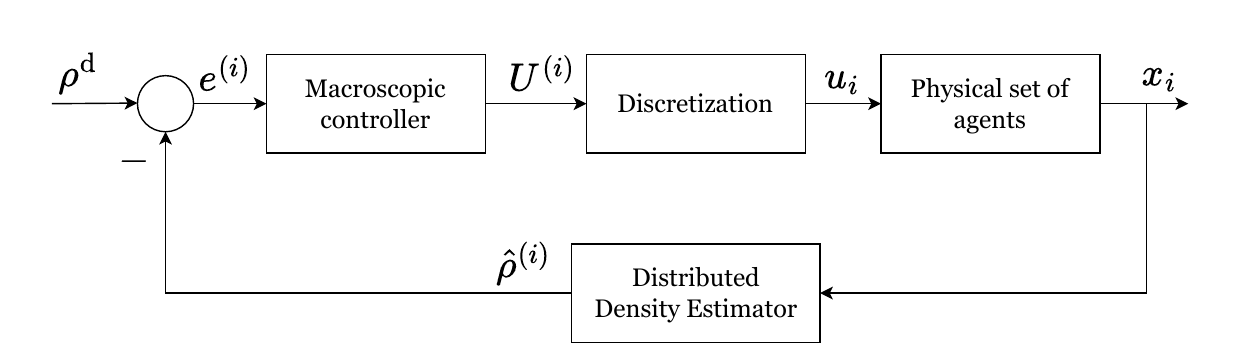}%
        \label{fig:dc_block_diagram}%
        }%
    \hfill%
    \subfloat[]{%
        \includegraphics[width=0.5\textwidth]{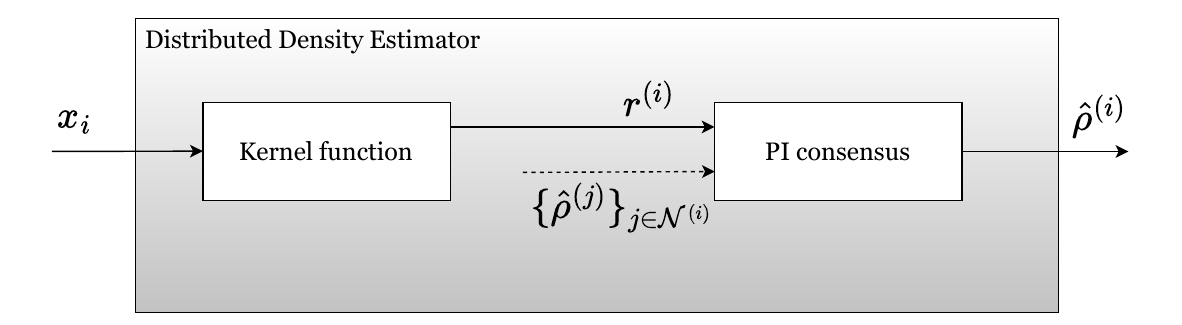}%
        \label{fig:estimator_detail}%
        }%
    \caption{(a) Block diagram: individual agents compute their own macroscopic control action $U^{(i)}$ depending on their local estimate of the density. (b) Detail of the distributed density estimator.}
    \label{fig:block_diagram}
\end{figure}

Fig. \ref{fig:block_diagram} illustrates the overall decentralized  control strategy which relies on a distributed density estimation algorithm.

\begin{figure*}
\centering
\subfloat[Centralized agents]{\label{fig:regulation_initial} \includegraphics[width=0.24\textwidth]{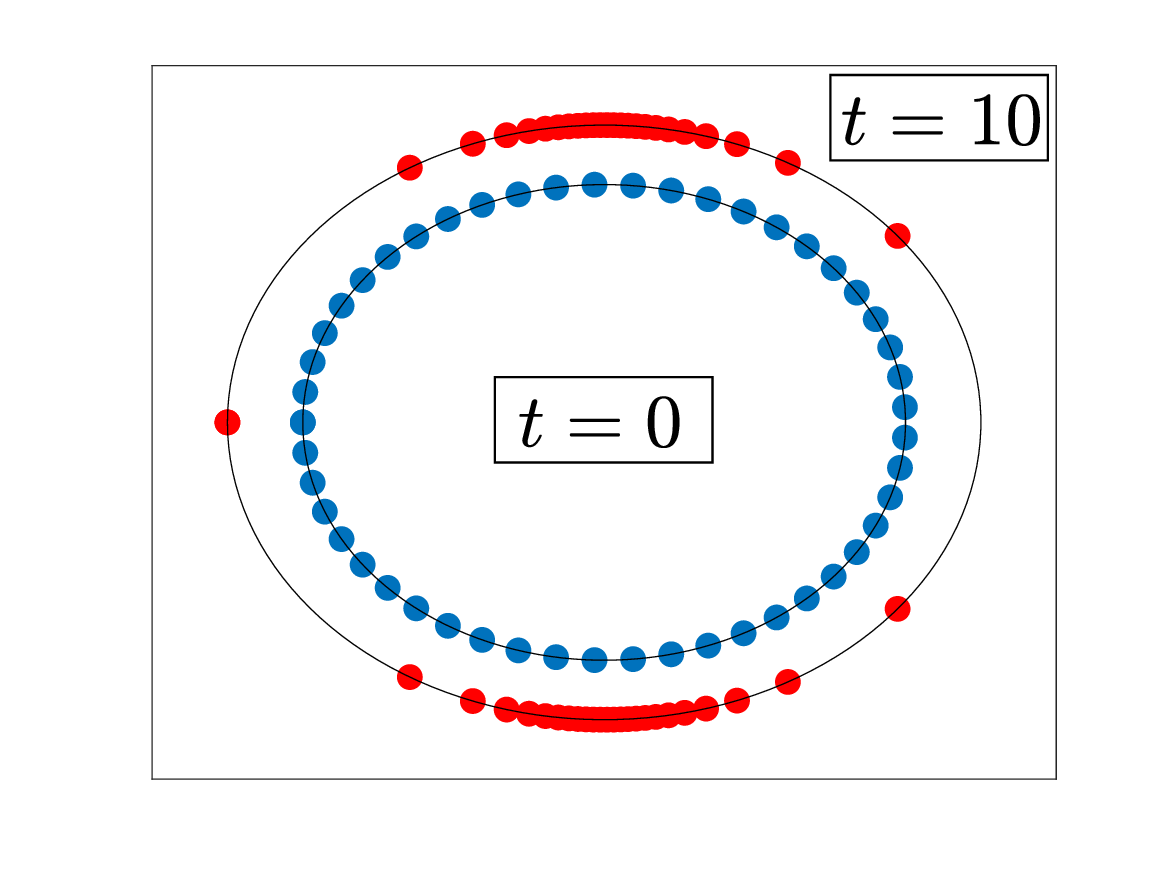}}
\hfill
\subfloat[Centralized densities]{\label{fig:regulation_final} \includegraphics[width=0.24\textwidth]{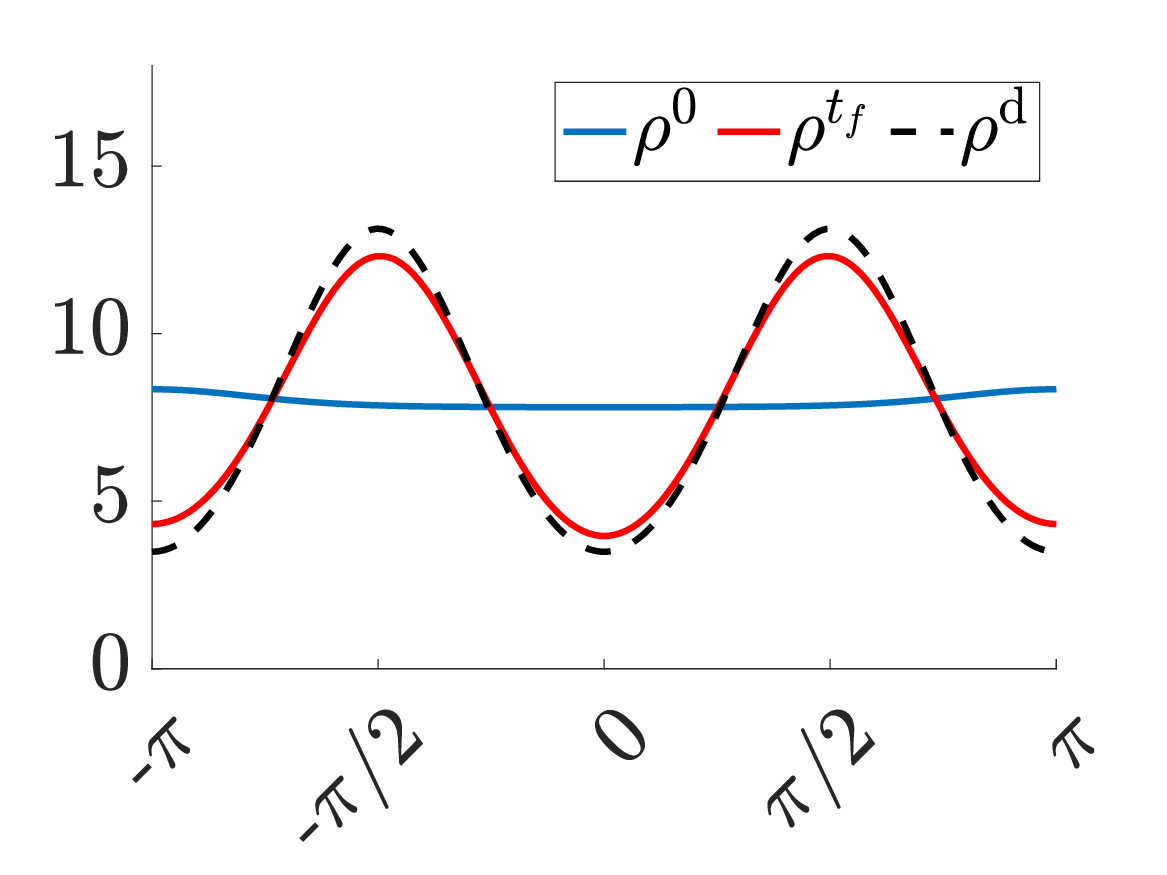}}%
\hfill
\subfloat[Decentralized agents]{\includegraphics[width=0.24\textwidth]{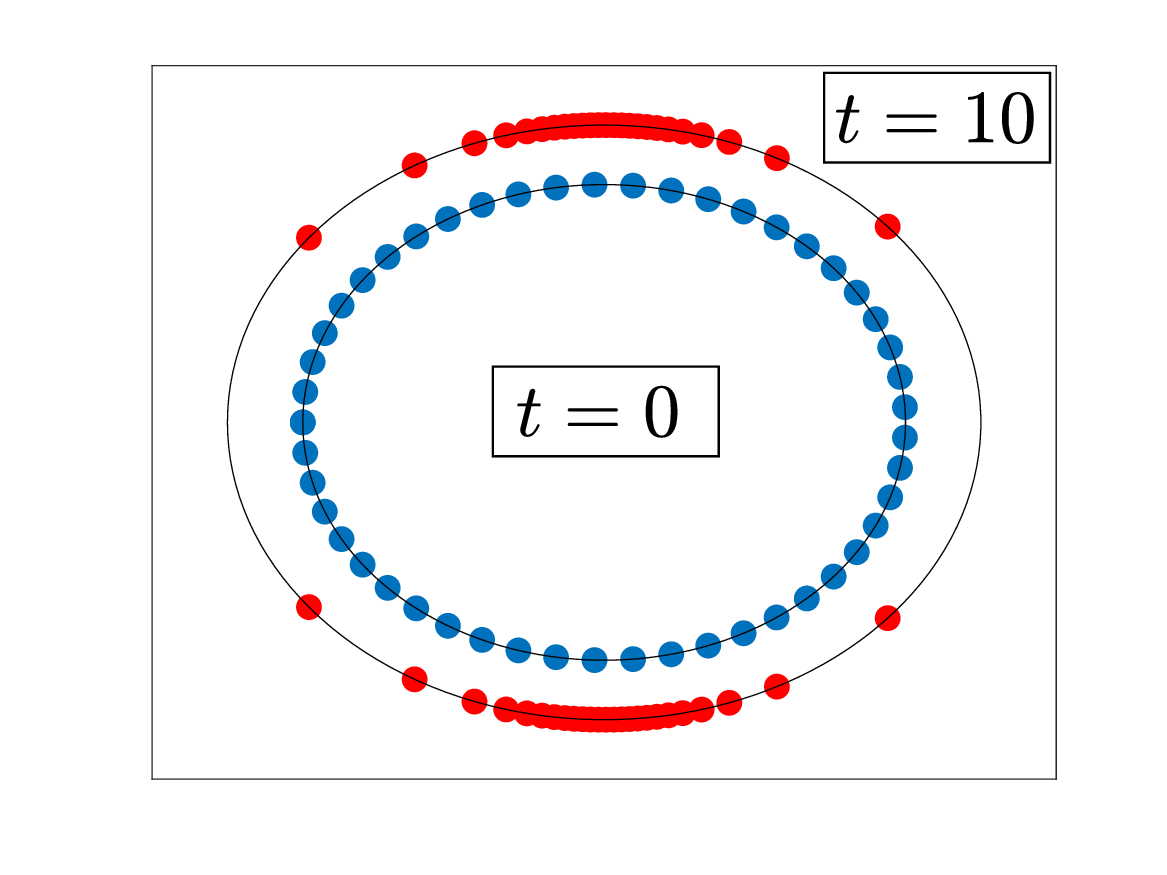}}%
\hfill
\subfloat[Decentralized densities]{\includegraphics[width=0.24\textwidth]{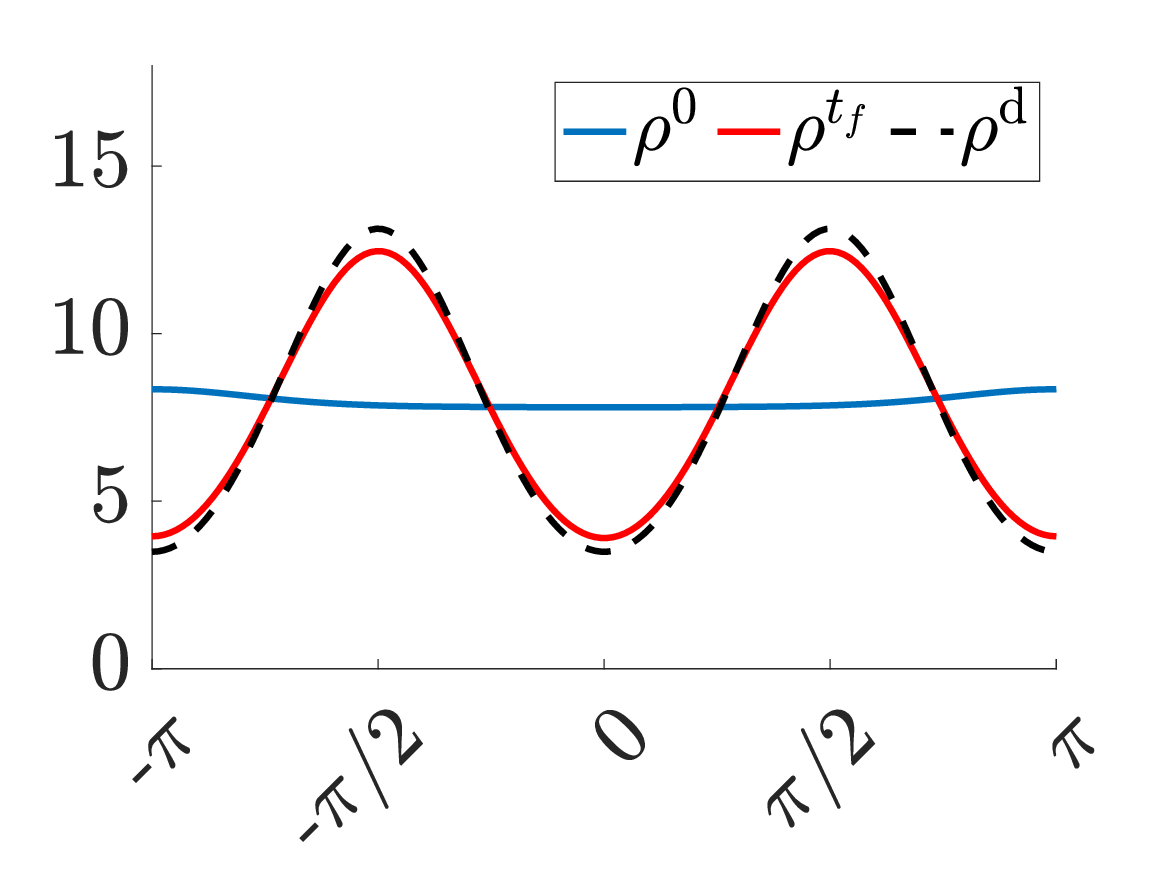}}%
\caption{Bimodal regulation trial. (a) Initial (inner circle) and final configuration (outer circle) of agents controlled with the centralized strategy, (b) initial (solid blue), final (solid orange) and desired (dashed black) densities; (c) initial (inner circle) and final (outer circle) configuration of agents controlled with the decentralized strategy, (d) initial (solid blue), final (solid orange) and desired (dashed black) densities.}
\label{fig:regulation}
\end{figure*}

\subsection{Distributed density estimation} \label{sec:density_estimation}
The group's density $\rho$ is computed from the agents' positions. Hence, to perform its estimation in a distributed fashion, we show how to design $c$ in \eqref{eq:density_estimator} such that $\hat{\rho}^{(i)}$ converges to $\rho$. Our methodology combines kernel density estimation (KDE) with PI consensus \cite{bernoff2011primer,kia2019tutorial}. 

In particular, within the field of kernel density estimation, density functions on $\Omega$ can be expressed as a summation of kernel functions
\begin{equation}\label{eq:rho_approx}
    \rho(x,t) = \sum_{i=1}^{N} K_h \left(x - x_i(t) \right),
\end{equation}
with $K_h:\Omega\times\mathbb{R}_{\geq 0} \to \mathbb{R}_{\geq 0}$, $h$ being a tunable smoothing parameter and
\begin{equation}
    \int_\Omega K_h(x,t) \, \mathrm{d} x = 1, \quad \forall t\geq 0,
\end{equation}

Up to a division by $N$, \eqref{eq:rho_approx} is the average of the kernel functions at each $x$. Therefore, we implement $c$ as a PI consensus algorithm to ensure that, if the communication topology $\mathcal{G}$ is strongly connected, $\hat{\rho}^{(i)}$ converges to the average in \eqref{eq:rho_approx} with a bounded steady-state error (see \cite{freeman2006stability}).

Hence, following \cite{freeman2006stability}, we select $c$ in \eqref{eq:density_estimator} so that
\begin{equation}
    \hat{\boldsymbol{\rho}}_t = -\alpha \left(\hat{\boldsymbol{\rho}} - \mathbf{r} \right) - \sigma_P\mathbf{L}\hat{\boldsymbol{\rho}} - \sigma_I\mathbf{L} \int_{0}^t \hat{\boldsymbol{\rho}} \, \mathrm{d} \tau.
    \label{eq:distroestimate}
\end{equation}
Here $\hat{\boldsymbol{\rho}} = [\hat{\rho}^{(1)} \dots \hat{\rho}^{(N)}]^\intercal$, $\alpha,\sigma_P$ and  $\sigma_I$ are positive tunable parameters, $\mathbf{L}$ is the Laplacian matrix associated to the communication network $\mathcal{G}$, and $\mathbf{r}$ is the stack vector containing the reference densities
\begin{equation}
    r^{(i)}(x,t) = N \, K_h \left(x - x_i(t) \right).
\end{equation}

The resulting decentralized continuification control strategy is then given by \eqref{eq:macroscopic_control} together with \eqref{eq:distroestimate}.

\section{APPLICATION TO MULTI-AGENT SYSTEMS} \label{sec:validation}
We validate our proposed strategy considering the multi-agent system \eqref{eq:agent-based_model} assuming agents' interactions are governed by a repulsive kernel, a common choice in the literature \cite{bernoff2011primer}. Following \cite{maffettone2024leader,di2024continuification}, we employ the periodization of a standard non-periodic kernel,
\begin{equation}
    f(x) = \mathrm{sgn}(x)\left[\exp\left(-\frac{|x|}{L}\right)\right],
\end{equation}
 where $L>0$ represents the interaction length scale. 

We set $L=\pi/4$ and study the problem of steering $N=50$ agents toward both static and time-varying configurations. We choose $K_\mathrm{p}=1$ and assume agents start evenly distributed on $\Omega$, resulting in an initial constant density  $\rho^0(x) = N/2\pi$. For the estimator, we empirically set $\alpha=1$, $\sigma_P = 5$, $\sigma_I = 5$, to guarantee a time-scale separation of at least one order of magnitude between the estimation and density error dynamics. We implement a 10 nearest neighbors (NN) communication graph, connecting each agent with its 10 nearest neighbors at $t=0$, which ensures topology connectivity.

For the PI estimator, we consider Von-Mises kernel functions as reference densities, 
\begin{equation}
    r^{(i)}(x,t) = N \, \frac{\exp \left(\frac{1}{h^2} \cos(x - x_i(t)) \right)}{2\pi I_0 \left(\frac{1}{h^2} \right)},
\end{equation}
where $h$ is empirically set to $0.7$.

We evaluate the performance of the control strategy using the following percentage density error:

\begin{equation}
    \bar{e}(t) := \frac{\| \rho^\mathrm{d}(\cdot,t) - \rho(\cdot,t) \|_2}{\max_t \| \rho^\mathrm{d}(\cdot,t) - \rho(\cdot,t) \|_2},
\end{equation}
where $\rho$ is the swarm's kernel density estimate centrally computed from agents' positions using \eqref{eq:rho_approx}.

To assess the estimation accuracy, we define the following average estimation error:
\begin{equation}
    \hat{e}(t) := \frac{1}{N}\sum_{i=1}^N\frac{\| \rho(\cdot,t) - \hat{\rho}^{(i)}(\cdot,t) \|_2}{\max_t \| \rho(\cdot,t) - \hat{\rho}^{(i)}(\cdot,t) \|_2}.
\end{equation}

In the remainder of this section, we examine several scenarios. Firstly, we analyze density regulation and tracking with agents communicating through a time-invariant strongly connected topology (10-agent NN graph). Then, we study the scenario in which agents communicate through a proximity network, relaxing the assumptiuon of a  strongly connected topology. Finally, we investigate how different communication graph topologies influence the convergence rate of density and estimation error.

\subsection{Regulation}\label{subsec:regulation}
We start with a regulation task where agents must achieve a constant target density $\rho^\mathrm{d}(x)$ set as the bimodal Von-Mises distribution with means $\mu_1$, $\mu_2$ and concentration coefficient $\kappa$,
\begin{equation}
    \rho^\mathrm{d}(x) = \frac{N}{4\pi I_0(\kappa)} \left[ \mathrm{e}^{\kappa \cos(x-\mu_1)} + \mathrm{e}^{\kappa \cos(x-\mu_2)} \right],
\end{equation}
where $I_0$ is the Bessel function of the first kind of order 0. Parameters are set to $\mu_1 = -\pi/2$, $\mu_2 = \pi/2$, $\kappa=2$. 

\begin{figure}
    \centering
    \begin{subfigure}{0.48\columnwidth}
        \centering
        \includegraphics[width=\linewidth]{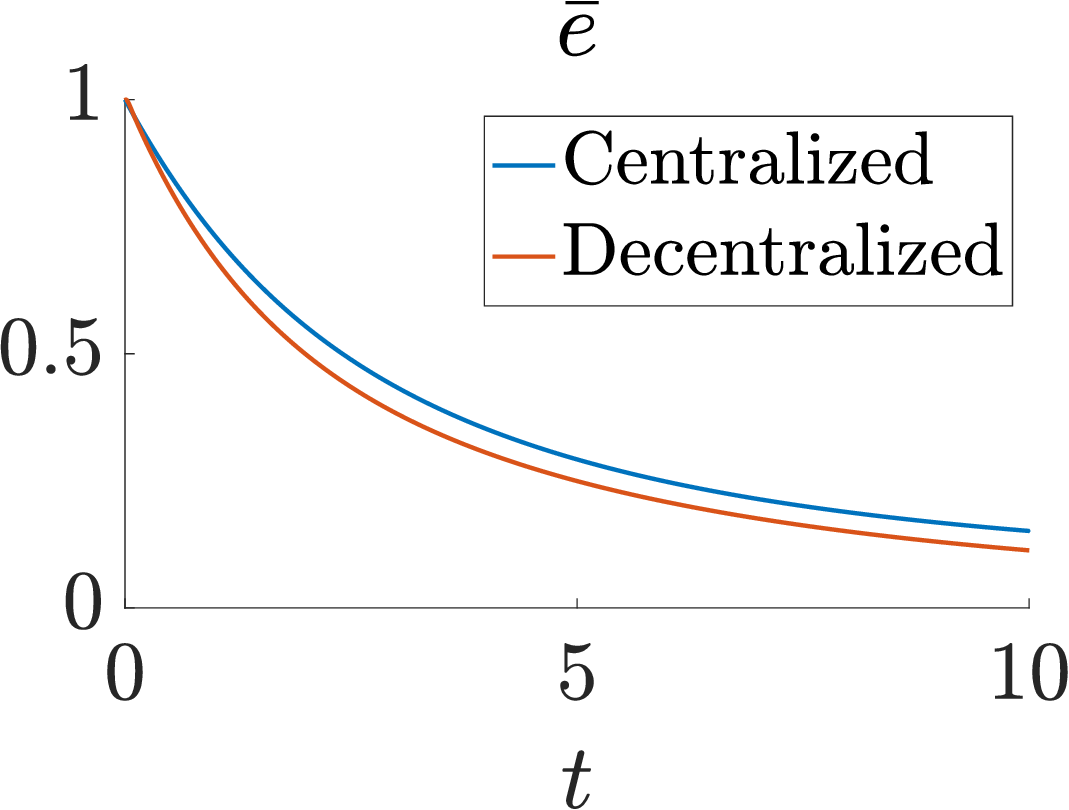}
        \caption{Density errors.}
        \label{fig:regulation_density_errors}
    \end{subfigure}
    \hfill
    \begin{subfigure}{0.48\columnwidth}
        \centering
        \includegraphics[width=\linewidth]{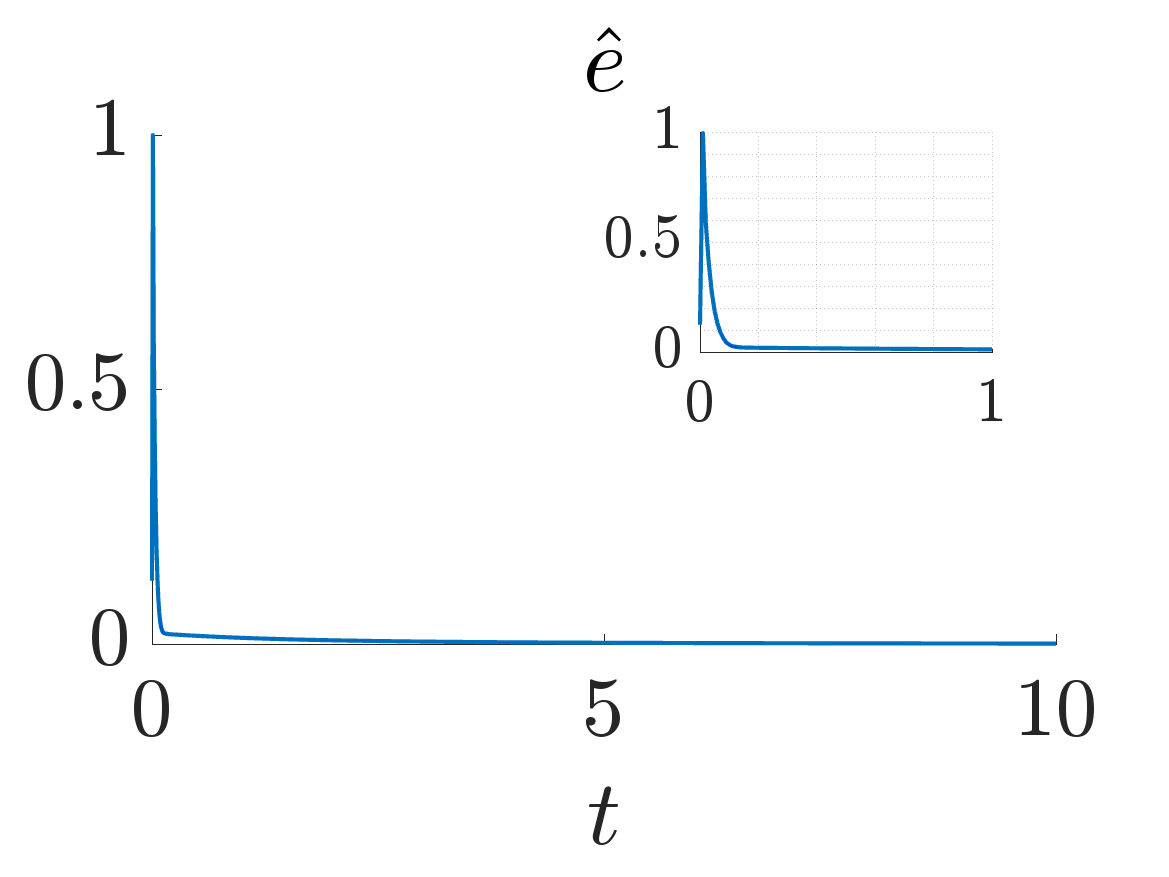}
        \caption{Estimation error.}
        \label{fig:regulation_estimation_errors}
    \end{subfigure}
    \caption{Bimodal regulation trial. (a) Comparison between density errors obtained with centralized and decentralized strategies; (b) estimation error with inset restricting the time interval to [0,1].}
    \label{fig:regulation_errors}
\end{figure}

In Fig. \ref{fig:regulation}, we compare the configurations achieved with the centralized and decentralized strategies, both in terms of agents' displacement and associated densities. Agents reach qualitatively similar steady-state configurations, as confirmed by Fig. \ref{fig:regulation_density_errors}, where we show the density error in time. Notice that, despite the results of Theorem \ref{th:macro_convergence} (which hold in the macroscopic setting for $N\to\infty$), a steady-state error is observed due to the discretization procedure described in Section \ref{subsec:discretization}. Fig. \ref{fig:regulation_estimation_errors} shows the evolution of the average estimation error.

\begin{figure}
    \centering
    \begin{subfigure}{0.48\columnwidth}
        \centering
        \includegraphics[width=\linewidth]{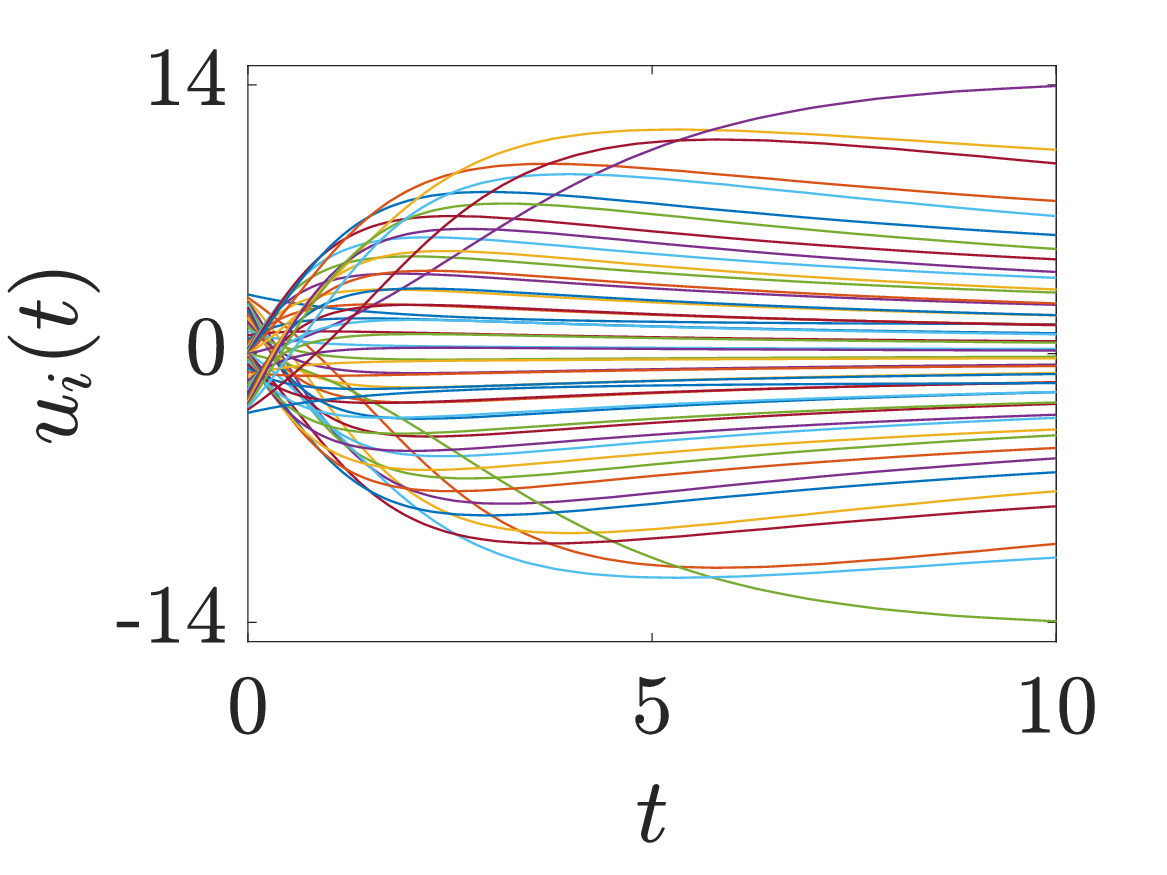}
        \caption{Centralized inputs.}
        \label{fig:regulation_centralized_inputs}
    \end{subfigure}
    \hfill
    \begin{subfigure}{0.48\columnwidth}
        \centering
        \includegraphics[width=\linewidth]{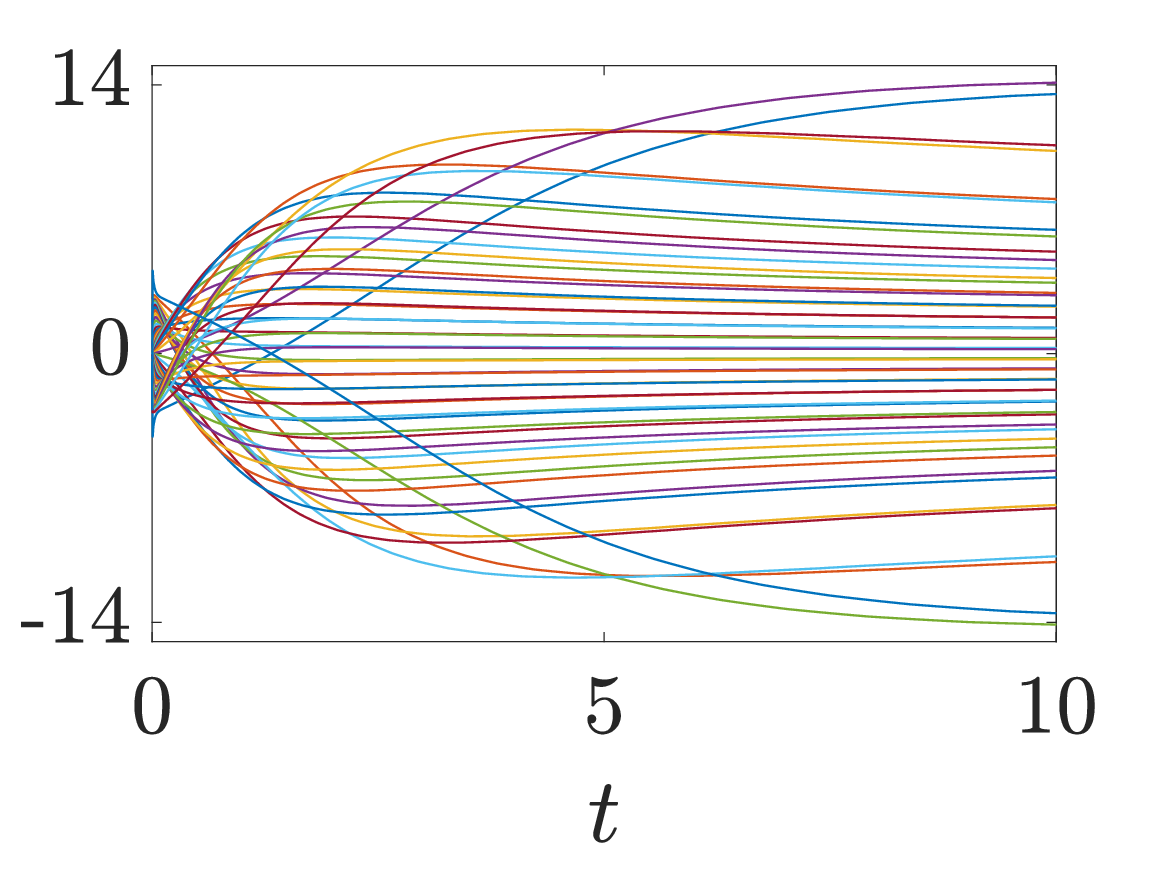}
        \caption{Decentralized inputs.}
        \label{fig:regulation_decentralized_inputs}
    \end{subfigure}
    \caption{Comparison between the microscopic control inputs of the regulation trial computed with the (a) centralized and (b) decentralized strategy.}
    \label{fig:control_inputs}
\end{figure}

The decentralized strategy exhibits faster error convergence (Fig. \ref{fig:regulation_density_errors}). This occurs because, during the transient phase of the estimator, the error computed by the controller is larger, leading to a stronger control action, as shown in Fig. \ref{fig:control_inputs}, where we compare the time evolution of the control inputs in the centralized and decentralized case. Asymptotically, the density error of the centralized solution converges to the same value as that of the decentralized solution.  

\subsection{Tracking}\label{subsec:tracking}
\begin{figure}
    \centering
    \begin{subfigure}{0.48\columnwidth}
        \centering
        \includegraphics[width=\linewidth]{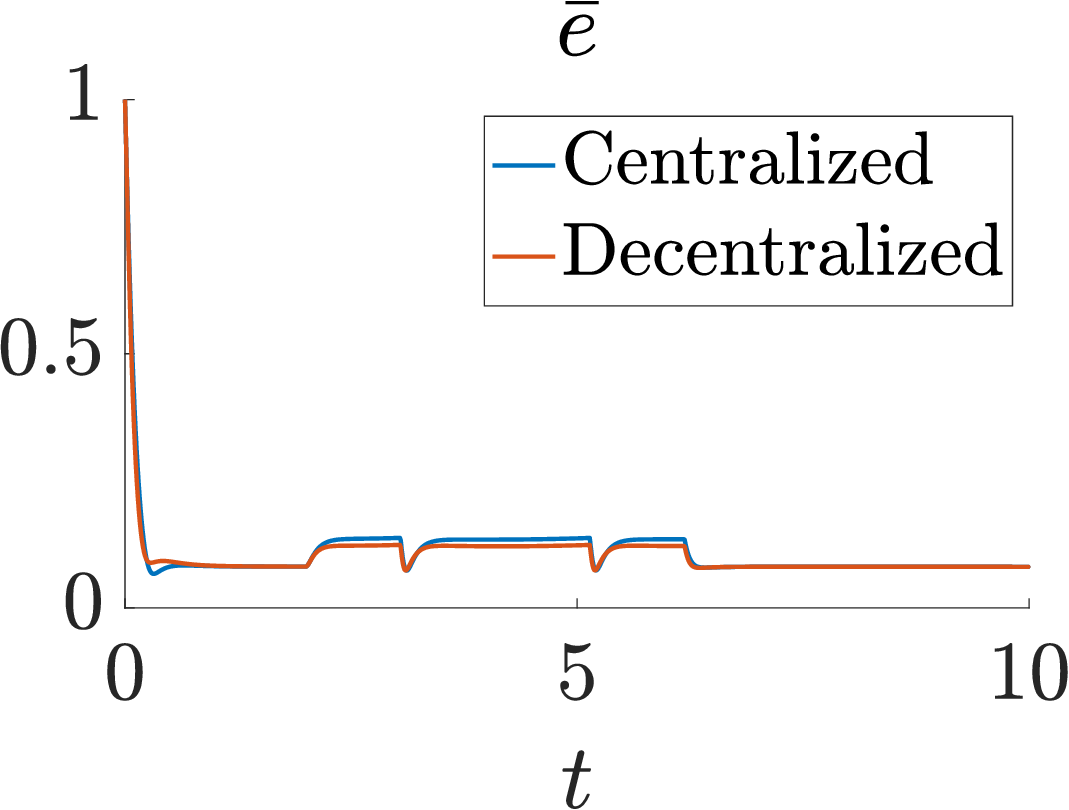}
        \caption{Density errors.}
        \label{fig:tracking_density_errors}
    \end{subfigure}
    \hfill
    \begin{subfigure}{0.48\columnwidth}
        \centering
        \includegraphics[width=\linewidth]{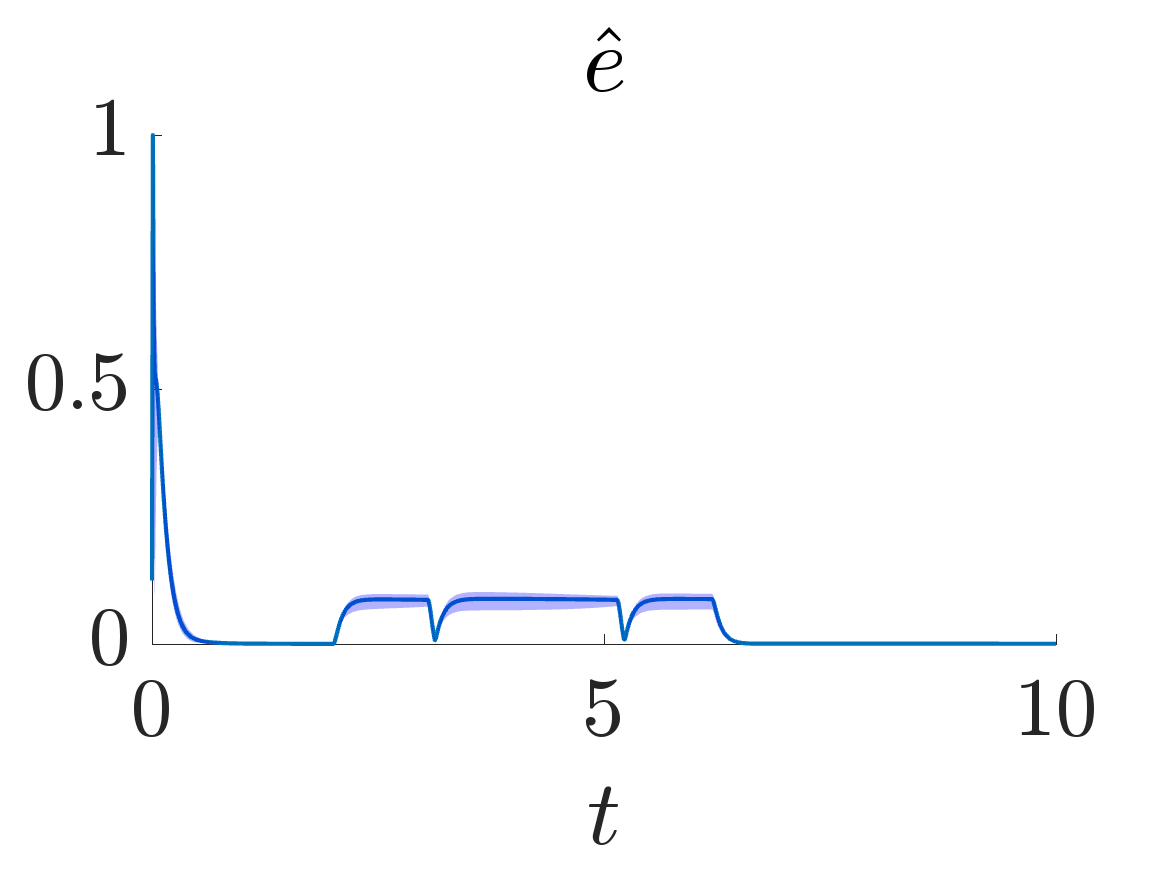}
        \caption{Estimation error.}
        \label{fig:tracking_estimation_errors}
    \end{subfigure}
    \caption{Tracking experiment. (a) Comparison between centralized and decentralized density errors; (b) estimation error (the shaded area represents the maximum and minimum estimation error exhibited by agents).}
    \label{fig:tracking}
\end{figure}

To assess the ability of the proposed strategy to track a time-varying reference, a mono-modal Von-Mises distribution with $\kappa = 1$ and a time-varying mean $\mu(t)$ is considered. The mean $\mu(t)$ is set to zero for $t\leq2$; then it increases at a rate of $\dot{\mu} = 1$ rad per time unit, until reaching $\pi/3$; subsequently, it decreases at a rate of $\dot{\mu} = -1$ until reaching $-\pi/3$; and finally, it increases again at rate $\dot{\mu} = 1$ until returning to zero. The density errors of the centralized and decentralized strategies are shown in Fig. \ref{fig:tracking_density_errors}, while the estimation errors are shown in Fig. \ref{fig:tracking_estimation_errors}. Both results confirm the effectiveness of the proposed strategy in estimating and tracking a time-varying density reference.

\subsection{Proximity communication network}\label{subsec:proximity}
\begin{figure}
    \centering
    \begin{subfigure}{0.48\columnwidth}
        \centering
        \includegraphics[width=\linewidth]{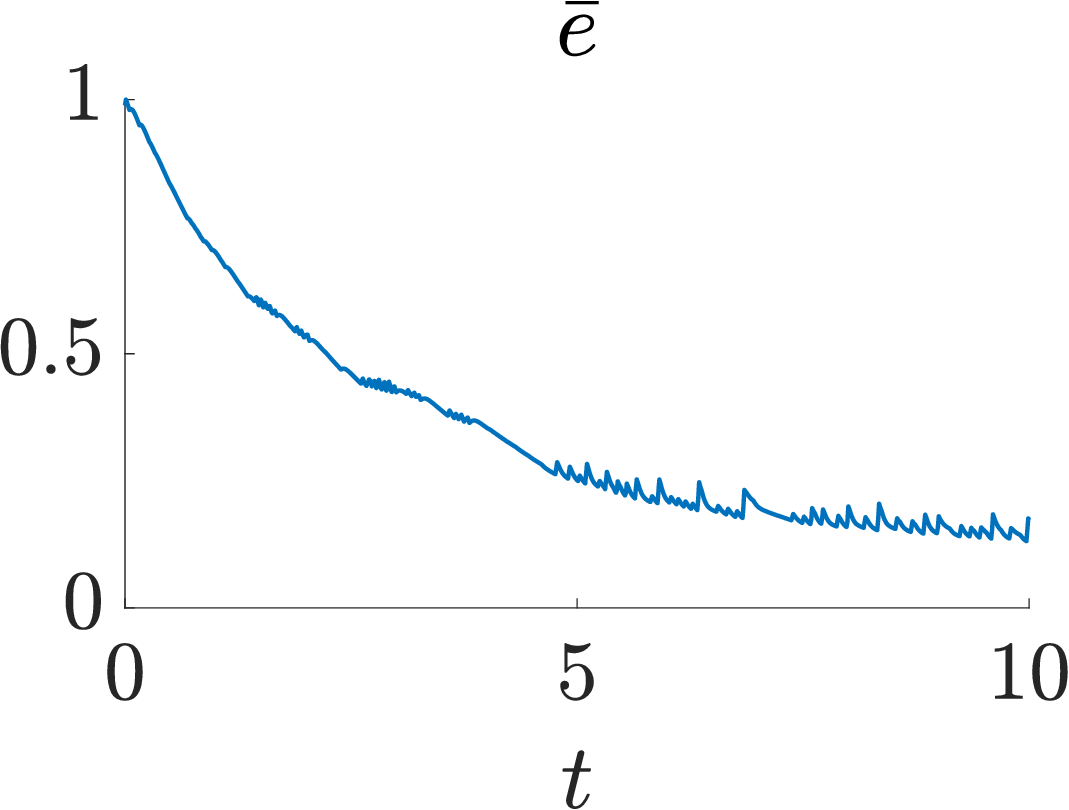}
        \caption{Density error.}
        \label{fig:proximity_density_error}
    \end{subfigure}
    \hfill
    \begin{subfigure}{0.48\columnwidth}
        \centering
        \includegraphics[width=\linewidth]{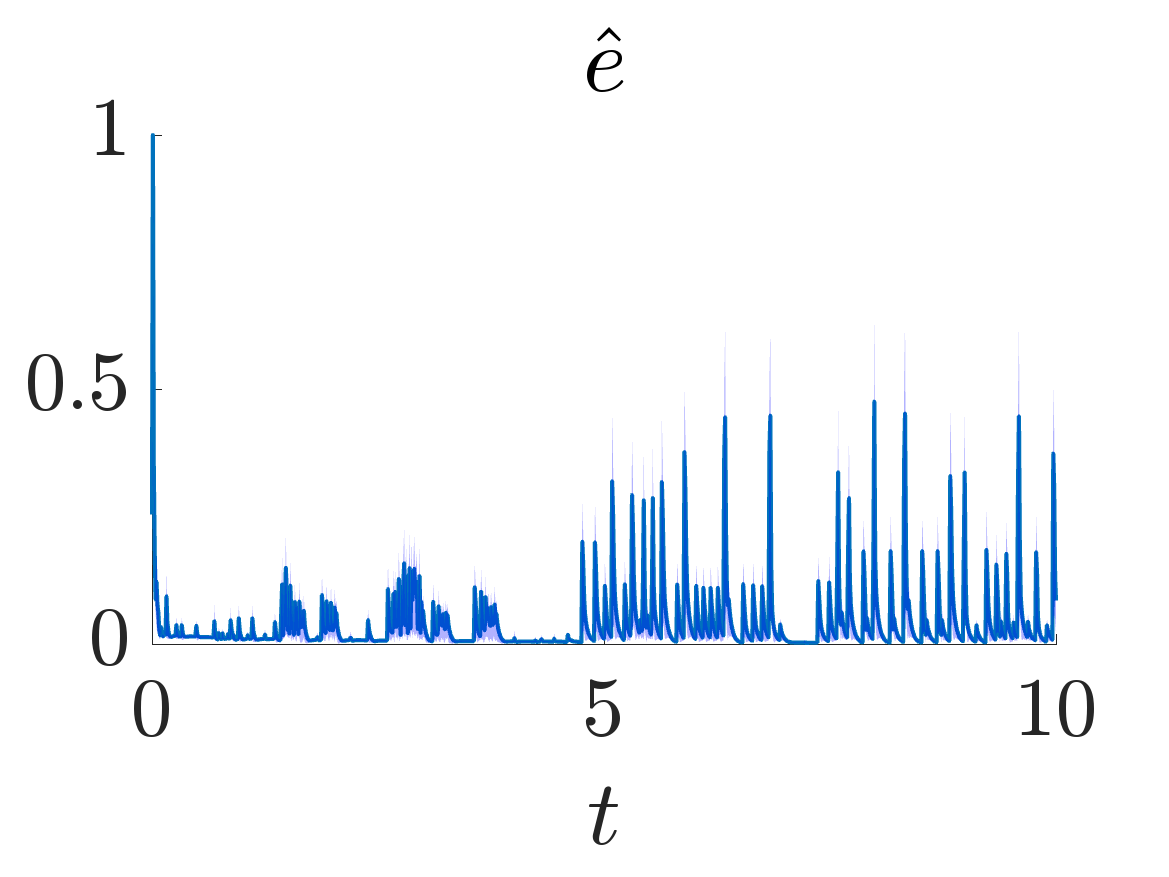}
        \caption{Estimation error.}
        \label{fig:proximity_estimation_error}
    \end{subfigure}
    \caption{Robustness assessment considering a proximity network as communication topology. (a) Density error; (b) estimation error (the shaded area represents the maximum and minimum estimation error exhibited by agents).}
    \label{fig:proximity}
\end{figure}

To evaluate the robustness of our approach to faults and perturbations of the communication network, we test it in a scenario where agents are able to communicate only when they enter each other's sensing areas. As a consequence, the communication network is not a time-invariant strongly connected graph but has a time-varying topology that, at times, can become disconnected. Agents are configured to exchange density estimates with neighbors within a neighborhood of radius $\varepsilon$ and the Laplacian matrix associated with this time-varying topology is used when implementing the PI estimator. Setting $\varepsilon = \pi/4$, we obtain the results shown in Fig. \ref{fig:proximity}. In comparison to the 10-nearest-neighbors topology (see Fig. \ref{fig:regulation_errors}), the density error in the proximity network converges less smoothly to zero, and the estimation errors display occasional short-lived spikes. Nevertheless, agents successfully achieve the desired steady-state configuration with a steady-state percentage error of approximately 0.25.


\subsection{Increasing number of neighbors}\label{subsec:NNnetworks}
\begin{figure}
    \centering
    \begin{subfigure}{0.48\columnwidth}
        \centering
        \includegraphics[width=\linewidth]{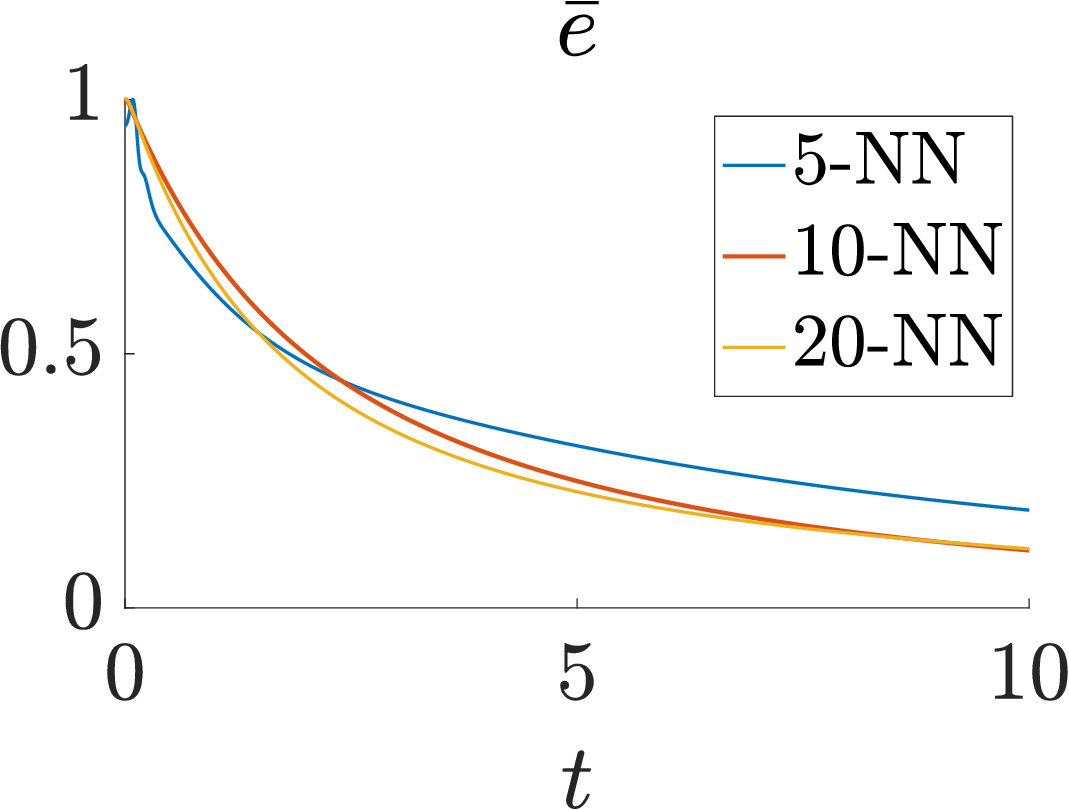}
        \caption{Density errors.}
        \label{fig:NN_density_errors}
    \end{subfigure}
    \hfill
    \begin{subfigure}{0.48\columnwidth}
        \centering
        \includegraphics[width=\linewidth]{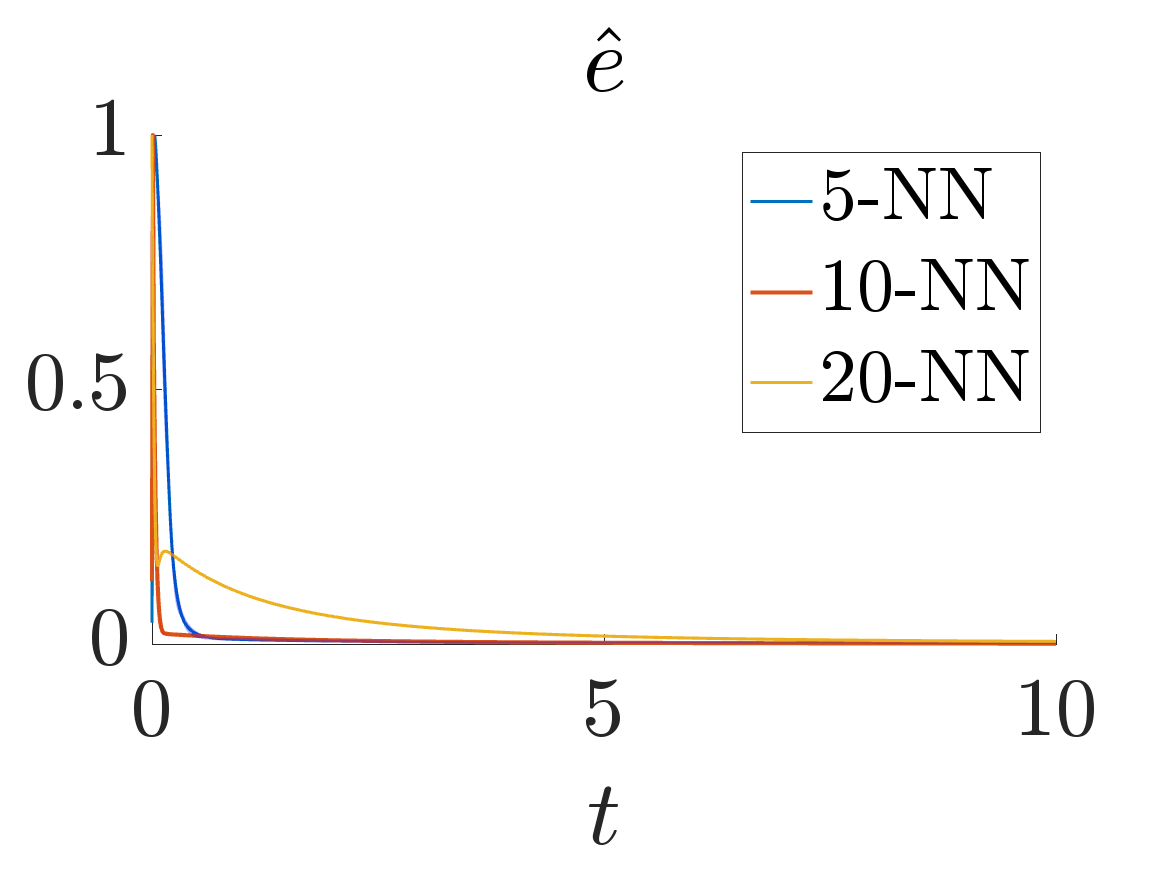}
        \caption{Estimation errors.}
        \label{fig:NN_estimation_errors}
    \end{subfigure}
    \caption{Evolution of the (a) density and (b) estimation errors with increasing number of nearest neighbors considered in the communication topology.}
    \label{fig:NNnetworks}
\end{figure}

Next, we assume agents can communicate on NN topologies with different characteristics. Communication topologies with 5, 10 and 20 NN are considered, and the evolution of the density and estimation errors is evaluated for the bimodal regulation problem described in Section \ref{subsec:regulation}. Results in Fig. \ref{fig:NNnetworks} show that density errors converge more rapidly to zero as the number of neighbors increases. Interestingly, in contrast, estimation errors exhibit faster convergence in the 10-NN topology. 

\section{CONCLUSIONS} \label{sec:conclusions}
We presented a distributed implementation of a continuification method for density control of large-scale multi-agent systems. The approach relaxes unrealistic assumptions about centralized knowledge of macroscopic observables by incorporating inter-agent communication for distributed density estimation. Numerical evidence convincingly demonstrates that when agents  exchange information through a consensus protocol over a strongly connected topology, effective density control can be achieved across various scenarios. While derived in a one-dimensional settings, our results can be extended to higher-dimensional domains following the approach presented in  \cite{maffettone2024mixed} for the centralized strategy.

Our work illustrates how control methodologies based on macroscopic descriptions of the system dynamics can be applied in realistic settings, where agents possess limited sensing capabilities and must operate without any centralized information.

Future work will address current limitations by analytically assessing how distributed density estimation affects the stability properties of the control strategy, drawing inspiration from the literature on PDE observers \cite{krstic2008boundary}. Additional investigations will examine how communication delays impact the estimation mechanism \cite{hasanzadeh2024distributed} and apply this distributed implementation for the leader-follower scenario discussed in \cite{maffettone2024leader}.


\balance
\bibliographystyle{IEEEtran}

\end{document}